\documentclass[]{aa}

\usepackage{mathptmx}

\usepackage[T1]{fontenc}
\usepackage{ae,aecompl}

\usepackage{graphicx}   
\usepackage{amsmath}    
\usepackage{amssymb}    
\usepackage{url}

\usepackage{tikz,xcolor,hyperref}

\definecolor{lime}{HTML}{A6CE39}
\DeclareRobustCommand{\orcidicon}{%
	\begin{tikzpicture}
	\draw[lime, fill=lime] (0,0) 
	circle [radius=0.16] 
	node[white] {{\fontfamily{qag}\selectfont \tiny ID}};
	\draw[white, fill=white] (-0.0625,0.095) 
	circle [radius=0.007];
	\end{tikzpicture}
	\hspace{-2mm}
}

\foreach \x in {A, ..., Z}{%
	\expandafter\xdef\csname orcid\x\endcsname{\noexpand\href{https://orcid.org/\csname orcidauthor\x\endcsname}{\noexpand\orcidicon}}
}

   \title{Drifting inwards in protoplanetary discs I \\Sticking of chondritic dust at increasing temperatures}
   
\titlerunning{Drifting inwards in protoplanetary discs I}

   \author{T. Bogdan
        \inst{1}\orcidA{}
        \and
        C. Pillich\inst{2}\orcidB{}
        \and
        J. Landers\inst{2}\orcidC{}
        \and
        H. Wende\inst{2}\orcidD{}
        \and
        G. Wurm\inst{1}\orcidE{}
}
\authorrunning{Bogdan et al.}

\institute{University of Duisburg-Essen, Faculty of Physics,
        Lotharstr. 1, 47057, Germany\\e-mail: tabea.bogdan@uni-due.de
        \and
        University of Duisburg-Essen, Faculty of Physics and Center for Nanointegration Duisburg-Essen (CENIDE), Lotharstr. 1, 47057 Duisburg, Germany\\
}

\date{}

\begin{document}

\abstract{
        Sticking properties rule the early phases of pebble growth in protoplanetary discs in which grains regularly travel from cold, water-rich regions to the warm inner part. This drift affects composition, grain size, morphology, and water content as grains experience ever higher temperatures. In this study we tempered chondritic dust under vacuum up to 1400\,K.
        Afterwards, we measured the splitting tensile strength of millimetre-sized dust aggregates. The deduced effective surface energy starts out as $\gamma_e = 0.07\,\rm J/m^2$. This value is dominated by abundant iron-oxides as measured by Mössbauer spectroscopy. Up to 1250\,K, $\gamma_e$ continuously decreases by up to a factor five. Olivines dominate at higher temperature. Beyond 1300\,K dust grains significantly grow in size. The $\gamma_e$ no longer decreases but the large grain size restricts the capability of growing aggregates. Beyond 1400\,K aggregation is no longer possible.
        Overall, under the conditions probed, the stability of dust pebbles would decrease towards the star. In view of a minimum aggregate size required to trigger drag instabilities it becomes increasingly harder to seed planetesimal formation closer to a star.}

\keywords{Planets and satellites: formation - Protoplanetary disks}

\maketitle

\section{Introduction}
\label{sec:introduction}

The primary trigger to terrestrial planet formation is growth of dust grains into larger, pebble size aggregates \citep{Blum2008, Johansen2014, Youdin2005}. This growth does not proceed in the same manner in all locations of protoplanetary discs. The most obvious example is the evolution of solids at the water snowline. The snowline separates an inner region with water being only present as vapour and an outer region with water ice \citep{Lecar2006ApJ}.  Water ice has been considered to be very sticky and to lead to dust traps \citep{Vericel2020} or the formation of highly porous planetesimals by sticking collisions alone \citep{Kataoka2013}. It has recently been shown however that the sticking properties of water ice strongly depend on the temperature \citep{Musiolik2019, Gundlach2018,Gaertner2017}. This might restrict the favourable conditions of icy planetesimals to a small region beyond the snowline, although sublimation and recondensation might play a role \citep{Ros2019}. 

However, this nicely shows that the temperature of the environment is important for collisional grain growth. For the inner regions many works considered silica or quartz as representative materials to describe sticking \citep{Blum2008,Dominik1997,Kimura2015,Steinpilz2019,Yamamato2014}. Temperature was less of an issue in this context in the past as $\rm SiO_2$ is not volatile at moderate temperature. A transition has regularly been considered only at the silicate ``snowline" (sublimation line) somewhere close to the star with temperatures beyond 1500\,K \citep{Morbidelli2016,Flock2019,McClure2013}. This picture of one stable silicate with constant sticking properties might be too simple in the context of planetesimal formation.

Inside of the water snowline dust consists of a number of minerals.
If we take chondrites as a representative material, we have olivines, pyroxenes, iron oxides, metallic iron, iron sulfide, and carbon-bearing minerals such as graphite just to mention the most abundant materials that can be assembled from the available chemical elements \citep{Braukmueller2018, Scott2014}. These minerals might have been subject to aqueous alteration depending on their exact location. Moderate temperature variations have already changed this assembly of minerals as is obvious from the classification scheme of meteorites with respect to thermal alteration \citep{Weisberg2006}. 
Therefore, grains at a location with 300\,K are different from grains at 1000\,K in more than one way, including composition, grain size, and morphology.

In this context, it is important to note that solids regularly drift inwards \citep{Weidenschilling1977}. This inward drifting brings minerals which are stable at low temperatures to much warmer regions and some mineral metamorphoses occur. How this influences the sticking properties of dust aggregates is subject of debate and not yet conclusive. In an initial study, \citet{deBeule2017} researched the tensile strength of JSC Mars simulant dust tempered in air, which is based on a new tensile strength measurement technique \citep{Musiolik2018}. These authors find some increase of sticking with increased temperature, although a decrease beyond 1000\,K would also be in agreement with the data. \citet{Demirci2017} carried out collision experiments in which basalt grains grew self-consistently to a maximum size (bouncing barrier). They found that sticking decreases for dust tempered beyond 1000\,K as aggregates only grew smaller then. This might also be important in view of recent calculations of planet formation in this hot region by \citet{Flock2019}. These authors find a particle trap located at 900\,K, which might just be a sweet spot for particle growth as it might be dry and sticky \citep{Steinpilz2019} but not yet be too hot for decreased sticking \citep{Demirci2017}.

In all these measurements grains were studied after they cooled down. Therefore, these measurements did not trace sticking at high temperature with all aspects of sticking at high temperatures, but only the influence that composition or grain size and shape might provide. Certainly, high temperature during a collision might have its own influence superimposed on this, that is from varying viscosity.
As an example, \citet{Bogdan2019} show that the coefficient of restitution for basalt grains starts to decrease at about 1000\,K. Also, \citet{Demirci2019b} find sticking probabilities to be increased for basalt in slow collisions for 875\,K upwards.

We continue along these pathways by studying systematically how grains in a chondritic sample, milled to dust and subject to different temperatures stick to each other. In contrast to previous studies, we temper under vacuum in this work. We also use standard splitting tensile strength measurements (Brazilian tests) and relate the results to changes in composition and morphology of the material provided by Mössbauer spectroscopy and grain size analysis.

This work includes a number of measurements, analyses, and interpretations. For clarity, we provide a detailed analysis of a data set that measures contact forces ``as they are'' in the laboratory. No special concern is taken with respect to surface water in this work. From a recent work by \citet{Steinpilz2019}, based on a suggestion by \citet{Kimura2015}, we know that surface water is important for the measurement of sticking properties. Especially in view of higher temperatures, silicates loose any surface water they would carry from the snowline and their sticking would initially increase as they dry.  We note that this modifies the findings of this paper. However, this process is more complex and that work is currently still ongoing. The principle results of this paper still hold as the underlying evolution of the stickiness of chondritic matter. 

\section{Laboratory experiments}
\label{sec:experiments}

\subsection{Chondritic dust}

We consider an initial mix of minerals drifting inwards in protoplanetary discs. In the laboratory, we simulated the drift by tempering the material at increasing temperatures. 
As far as the sample is concerned, this is an empirical study. We do not consider in detail, which minerals would change into new minerals depending on the detailed compositions from first principles. We rather intend to sample the early solar nebula in its origin as well as possible. 
Therefore, we used chondritic material in this study.  We acquired a meteorite (Sayh al Uhaymir 001), which is classified as L4/5 chondrite. This is different from the materials used in earlier studies \citep{Demirci2017, deBeule2017}. We relied on the vendor for the classification in this work. Cutting the stone, it shows chondrules and metal inclusions in parts, but otherwise we did not do any further analysis to verify the chondrite type. In the context of this study, this is of minor importance, especially as we traced the composition upon heating in detail (see below).

 
Tempering this chondritic material might have two effects. First, the internal mineral phases might change into new mineral phases. The surface of minerals subject to a gaseous environment or vacuum (or inert gas) might change in a different way \citep{Braukmueller2018}. It is therefore important if the whole stone is tempered or if the stone is ground first and the dust is then tempered. In the latter, there is a large effective surface that is subjected to the environment.
In view of our interest in dust evolution, the meteorite has to be ground before being subject
to higher temperatures. 

Second, the grain size is also important as sticking properties strongly depend on it. Tempering might change the effective grain sizes.
We do not aim at producing exactly a grain size that might have been present in the solar nebula, which is hardly possible.
However, in the comparison of samples tempered differently, the initial grain size has to be at least the same.
A further requirement is that the particles have to be small enough to be sticky at all to allow tensile strength measurements using a Brazilian test.  Grains are therefore in the micrometer range.
We developed a sequence of milling the material in the same way and used the same batch of ground material for a series of tempering. 

\begin{figure}
	\includegraphics[width=\columnwidth]{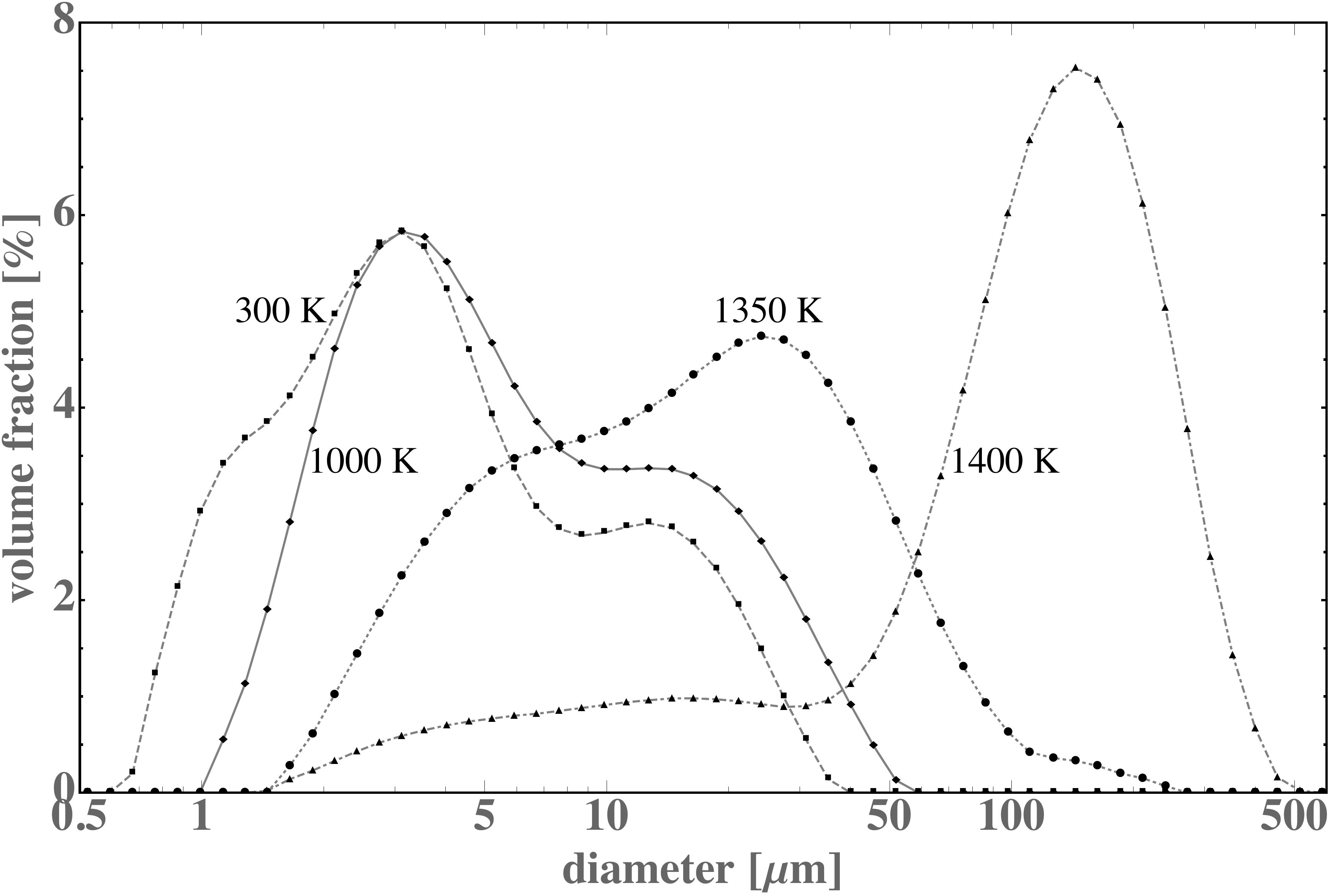}
	\caption{\label{fig.grainsizeexample}Grain size distribution (volume fraction) for tempered samples. Two examples for sizes below 1250\,K (300\,K and 1000\,K)  are similar. At higher temperature the grain size shifts to larger sizes. The sample at 1400\,K is dominated by large grains of 142\,$\rm \mu m$.}
\end{figure}

\subsection{Measurement sequence}

In total we only had about 20\,g of the meteorite available. As tempering changes the material, a 
certain sequence for the different measurements had to be set up.
A part of the heated material at each temperature was used for Mössbauer spectroscopy. This sample is not destroyed during this procedure and is afterwards also used for measuring the size distribution in the sample. Tensile strength measurements were carried out with the main
part of the material. These measurements are representative of the temperature at which the material was tempered.
Afterwards, the material was subjected to a higher temperature. Owing to this alteration, no more measurements at lower temperatures are possible unless a new batch of material is prepared. 

\subsection{Grain size evolution with temperature}

Fig.\,\ref{fig.grainsizeexample} shows examples of the (volume) grain size distribution measured with a commercial instrument (Mastersizer 3000). The instrument determines grain sizes by light scattering and Mie analysis, thus giving equivalent diameters of spherical grains based on optical properties.

\begin{figure}
        \includegraphics[width=\columnwidth]{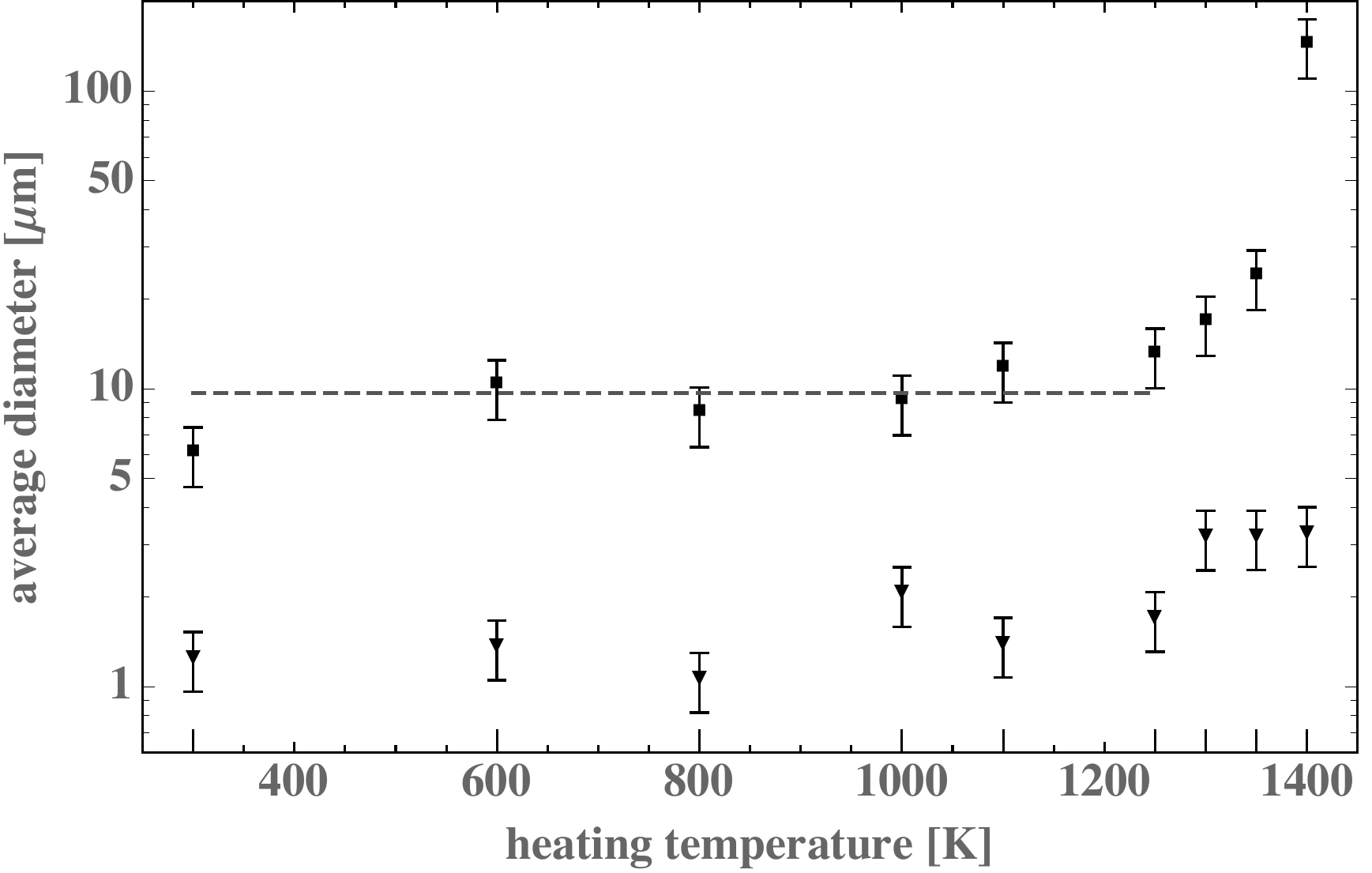}
        \caption{\label{fig.grainsizeevolution}Grain size evolution with temperature. The volume-averaged grain sizes (squares) and the number averaged grain sizes (triangles) are shown. A line indicates the mean diameter of $9.7\pm2.5\, \mu\mathrm{m}$ for the temperature range from 300\,K to 1250\,K based on the volume-averaged grain size.}
\end{figure}

Optical constants enter into these calculations and we used $n = 1.518 + 0.001 \rm i$ as refraction and absorption index, respectively, throughout all measurements. This has to be kept in mind as small differences during the evolution of the sample might be related to small changes in optical constants and to changes in grain shape. 
However, the samples only show a significant increase in grain size at temperatures beyond 1250\,K.
Fig.\,\ref{fig.grainsizeevolution} shows the evolution of the average grain size with temperature.  

Up to 1250\,K the size can be considered to be constant with a mean diameter of $9.7\pm2.5\, \mu\mathrm{m}$. The diameter increases to $16.6\,\mu\mathrm{m}$, $23.7\,\mu\mathrm{m,}$ and $142.1\,\mu\mathrm{m}$ for 1300\,K, 1350\,K, and 1400\,K, respectively. 
We used the volume-averaged grain size for the calculations below as we consider this to be the relevant quantity in our context of the Brazilian test (see below).  Not used but noted for completion, the number averaged grain size is a factor of seven smaller. 

\subsection{Bulk density evolution}

\begin{figure}
        \includegraphics[width=\columnwidth]{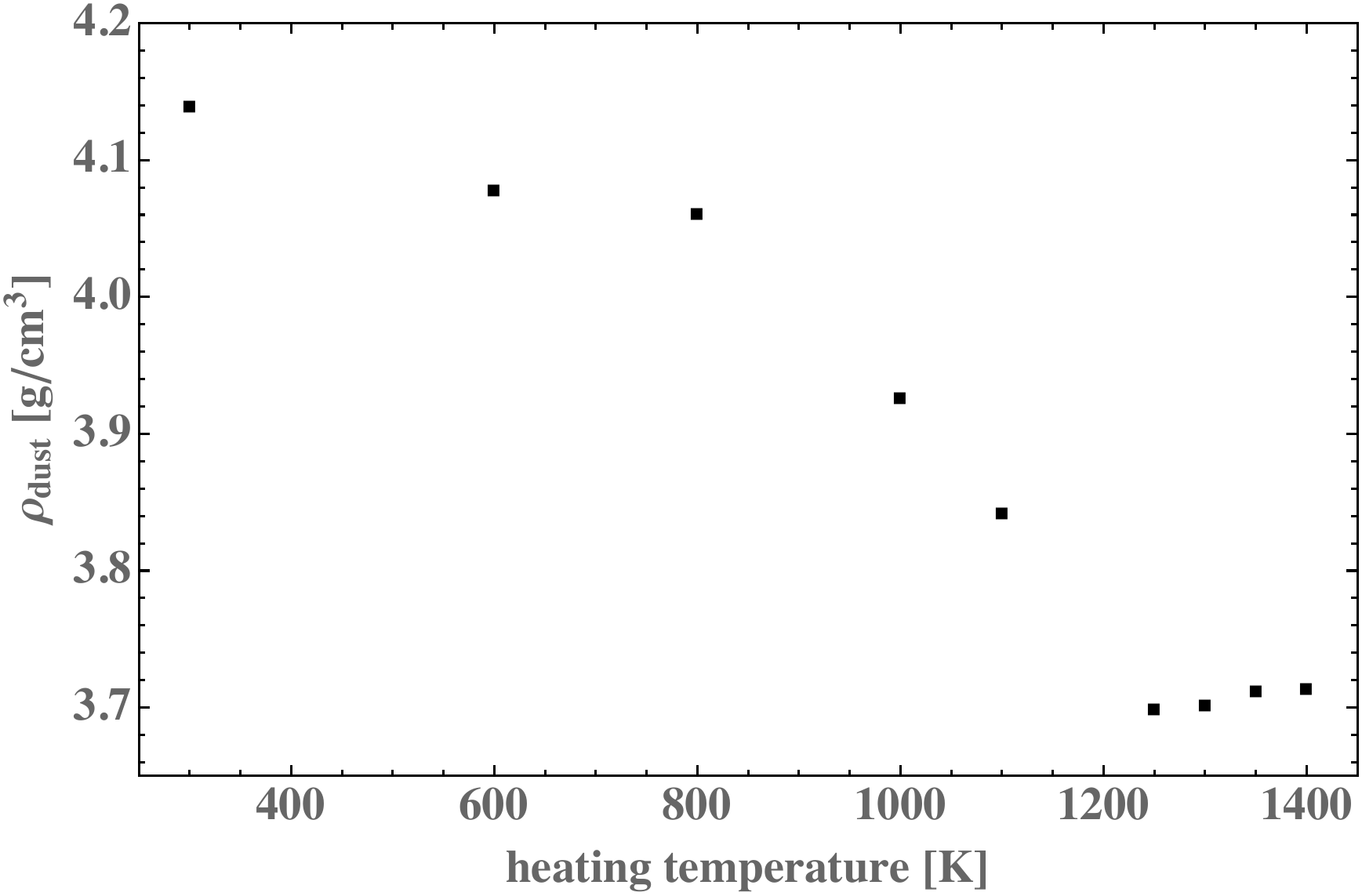}
        \caption{\label{fig.bulkdensity} Evolution of the bulk density of the material calculated from the composition determined in Mössbauer measurements.}
\end{figure}

As seen below, the Brazilian test strongly depends on the volume filling factor of a cylindrical dust aggregate. While the total volume and mass of an aggregate can be determined to high accuracy, the volume filling factor also depends on the density of the dust grains, which changes with change in composition upon tempering.
The meteorite as uncut stone has a density of 3.44\,$\rm g/cm^3$ based on volume measurements with a pycnometer. As chondrites have a certain porosity, the average grains of a powder after milling might have a somewhat higher density. 
As approximation for the powder we used the composition calculated from the Mössbauer measurements (see below). We note that this only includes ferric phases. However, it traces a slight decrease in density with increasing temperature as seen in Fig.\,\ref{fig.bulkdensity} and we used these values for the samples tempered at the different temperatures, respectively. 

\subsection{Tempering}


Exemplary studies (not shown), where the sample was kept at similar temperatures for $\mathrm{100 \, h}$ instead, showed that the duration has no significant influence on the minerals formed with a moderate decrease of phase transition temperatures by approximately $\mathrm{100 - 150 \, K,}$ as compared to dust heated for one hour we focus on in this study. We therefore consider a heating time of $\mathrm{1 \, h}$ to be a good representation of possible processes in a protoplanetary disc. As dust components transform into other minerals with higher condensation temperatures, we also do not expect changes in grain size for longer heating duration at least up to 1250\,K. Up to this point, we are still below the melting point of the abundant olivine. 

In detail, fayalite, as an iron-rich end member of olivine, has a melting temperature of 1200\,K at low pressure \citep{Hsu1967}. The melting temperature increases with increasing Mg-content of the olivine up to 1900\,K for forsterite \citep{Ohtani1981}.
The large grain size at 1400\,K suggests that the melting point of the olivines of the given composition is reached at this temperature, matching the intermediate Fe content in silicates as determined in Mössbauer spectroscopy (see below). 

\subsection{Mössbauer spectroscopy measurements}
Mössbauer spectra were recorded using a constant acceleration set-up and a $\mathrm{^{57}}$Co(Rh) source. To obtain spectra in a wide temperature range including magnetic phase transitions of the iron-bearing minerals, a liquid helium bath cryostat (CryoVac) was utilised. The temperature range of $\mathrm{4.3 \, K}$ to $\mathrm{60 \, K}$ was studied in detail to monitor magnetic phase transitions of iron-bearing silicates and the transition to the  superparamagnetic (SPM) behaviour of nanophase iron oxide. Mössbauer spectra were theoretically reproduced using the Pi program package developed by U. von Hörsten (\citet{Hoersten}). All values for the isomer shift $\mathrm{\delta}$ are given relative to that of $\mathrm{\alpha}$-Fe at room temperature ($\mathrm{\delta_{\alpha-Fe}=-0.107 \, mm/s}$) (\citet{Barb1980}).

\subsection{Cylinder preparation}

The procedure described above provides the base treatment of grinding and tempering of the meteorite. For the tensile strength measurements by means of the Brazilian test, this powder has to become a cylinder of dust. 
For a measurement, the dust is therefore pressed in cylindrical moulds of $7.99 \pm 0.05$\,mm diameter and $6.89 \pm 0.49$\,mm height. Varying the compression force provides different average densities or volume filling factors. This process is carried out manually because, depending on the evolution of the contact forces, more or less pressure is needed to provide a stable cylinder. The cylinders later break as supposed to in a Brazilian test (see below), and we consider the inner structure after compression to be homogeneous.

The mass of each cylinder was measured using a microscale. The mass was divided by the volume of the cylinder to get the average cylinder density. The ratio of the cylinder density to the bulk density of the dust estimated above gives the volume filling factor $\Phi$ 

\begin{equation}
\Phi = \frac{V_{dust}}{V_{cylinder}} = \frac{\rho_{cylinder}}{\rho_{dust}}.
\label{eq:filling_factor}
\end{equation}

\subsection{Tensile strength experiments}

A basic quantity for aggregation is the force $F_b$ needed to break one contact between two grains. 
For two (hard) spherical particles with reduced radius $R$ that are of same material and have a surface energy $\gamma,$ this can be calculated as \citep{Johnson1971}

\begin{equation}
F_b =  3 \pi \gamma R.
\label{jkr}
\end{equation}

This simple equation clearly visualises the two basic parameters important in this context, the surface energy and the particle size. It has to be noted that the sticking force increases linearly with grain size.
So in a static sense, two large grains are more sticky than two small grains. However, in a given cross section of an aggregate of tightly packed grains the number of contacts depends on $1/R^2$. Therefore, aggregates composed of smaller grains, in general, are more stable. 

Contact forces between two dust grains can in principle be measured directly, for example as done by \citet{Heim1999}, with atomic force microscopy for spherical grains.
However, eq. \ref{jkr} is an idealisation of any natural dust sample. Dust always comes in polydisperse samples to some degree with some variation in particle size. Grains are usually not spherical but irregular. Therefore, the contact force depends on the specific contact in an assembly and the local surface curvature. In addition, in a mix of minerals the surface energy also varies depending on the mineral species. It would therefore be a tedious task to measure sticking forces between individual grains of the chondritic dust and elaborate on its consequences.

To immediately get the important average value, we measured the splitting tensile strength $\sigma$ of dust aggregates here. 
The tensile strength of dust aggregates is connected to the contact force by Rumpf's equation \citep{Rumpf1970}, that is
\begin{equation}
\sigma = \frac{9 \Phi N}{8 \pi d^2} F.
\label{rumpf}
\end{equation}
In this equation, $N$ is the coordination number (number of touching neighbours),  $d$ is the diameter of an individual grain, and $F$ is the relevant average force to break a contact. This force might be the sticking force of eq. \ref{jkr} but it might also be rolling or sliding forces, whichever is weaker. All forces depend on the surface energy $\gamma$ however \citep{Dominik1997, Kimura2015, Musiolik2019}.

\begin{figure}
        \includegraphics[width=\columnwidth]{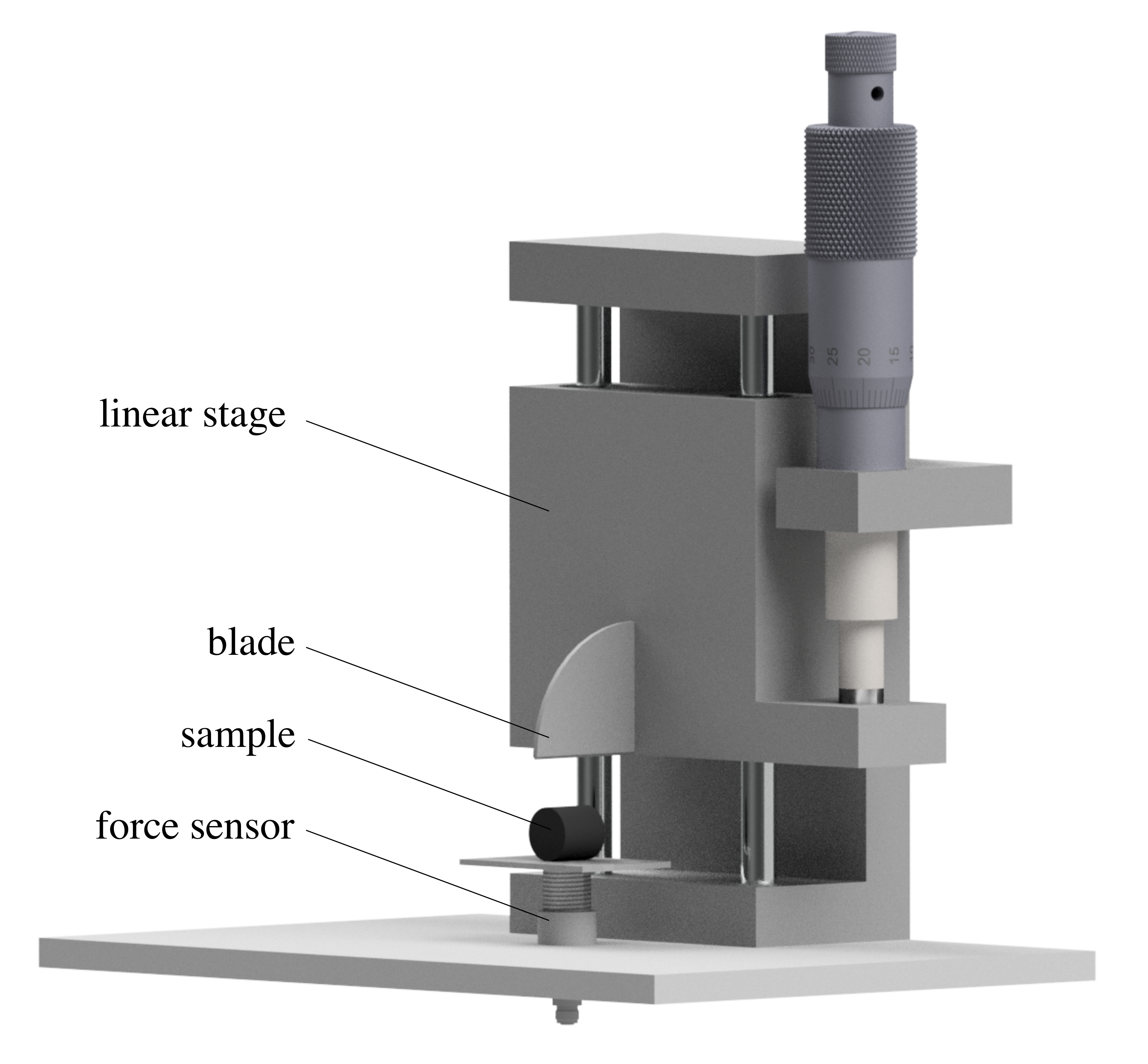}
        \caption{\label{fig.setup} Experimental set-up. By driving down the linear stage, the blade applies a force onto the cylindric sample until it breaks. The peak force is measured by a force sensor. }
\end{figure}

\begin{figure}
	\includegraphics[width=\columnwidth]{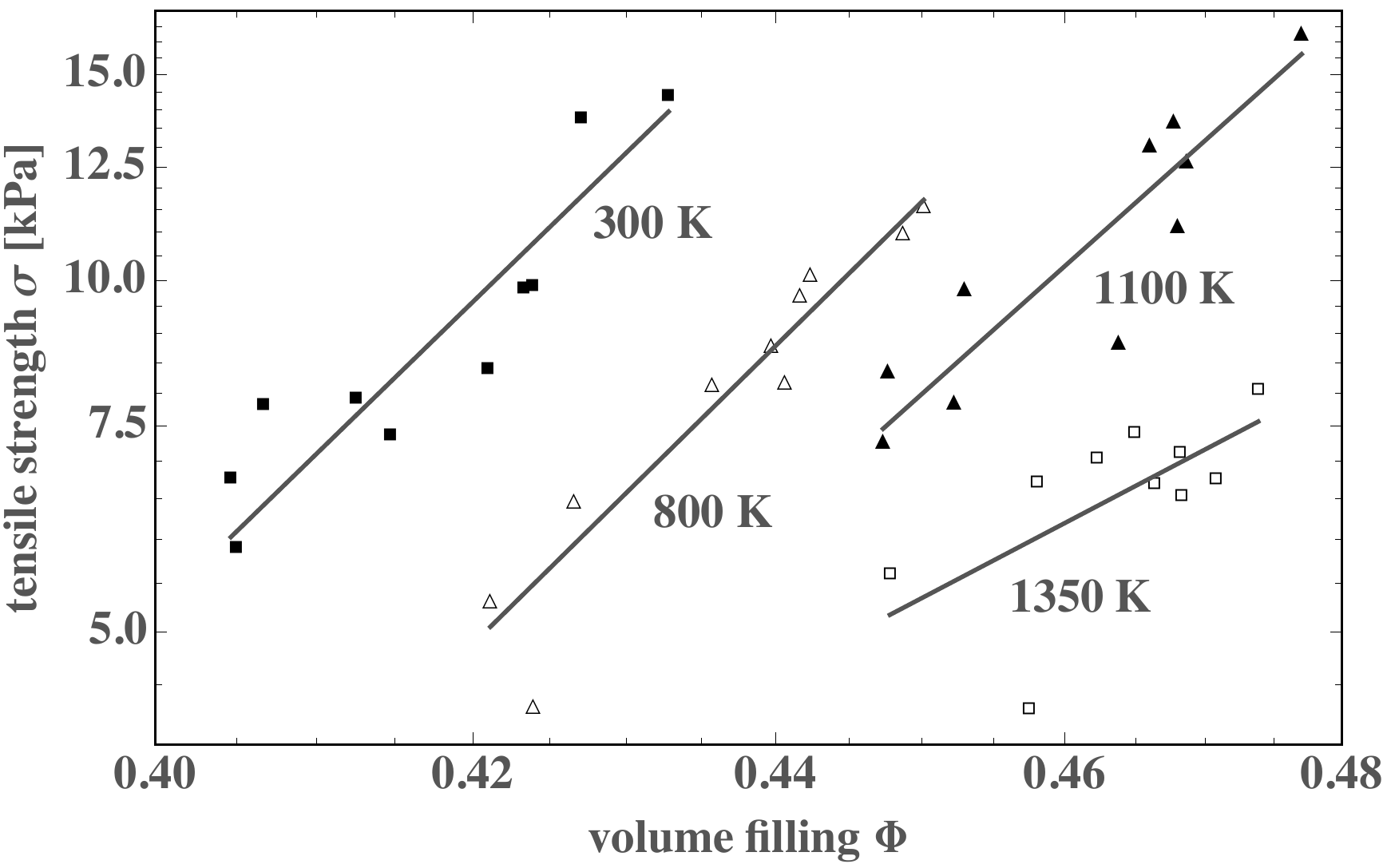}
	\caption{\label{fig.twetfit} Tensile strength $\mathrm{\sigma}$ over volume filling factor $\mathrm{\Phi}$ for tempered samples. Power laws are fitted to the data according to eq. \ref{andexer}. }
\end{figure}

To measure the tensile strength, we used a well-established method known as the Brazilian test. Examples where this technique has been applied in the context of matter in protoplanetary discs are \citet{Meisner2012}, \citet{Gundlach2018}, and \citet{Steinpilz2019}.
A sketch of the set-up we built for this purpose is shown in Fig.\,\ref{fig.setup}.  Pressure is applied onto the mantle face of each cylindrical sample until it breaks into halves. The peak force $F_t$ applied is measured by a force sensor and can be translated into the tensile strength 
\begin{equation}
    \sigma = 2F_t /(\pi d_c L)
,\end{equation} 
with diameter $d_c$ and length $L$ of the respective cylinder. The experiment is carried out at ambient conditions, that is room temperature and laboratory humidity. 
Fig.\,\ref{fig.twetfit} shows the tensile strength measurements with tensile strength over filling factor for different temperatures.

\section{Data analysis and interpretation}

\subsection{Tensile strength and volume filling factor}

To reduce the tensile strength data shown above, we start from Rumpf's equation (eq. \ref{rumpf}). The volume filling factor $\Phi$ is included explicitly but also implicitly in the number of contacts $N$. It has to be considered that also the average Force $F$ per contact depends on the filling factor for irregular grains. As the material is compressed, grains shift, contacts rearrange, and small force contacts with small asperities might be traded for larger contact areas and higher forces. 
Therefore, eq. \ref{rumpf} is written as
\begin{equation}
\sigma = \left(\frac{9}{8 \pi} \right) \left( \frac{1}{d^2} \right) \left( \Phi N_\Phi F_\Phi\right).
\label{rumpf2}
\end{equation}
In a first approach, for aggregates that are not too extremely compact and in agreement to Fig.\,\ref{fig.twetfit}, the total $\Phi$-dependence of $\sigma$ in each data set can be described by a power law 
\begin{equation}
    \sigma = \sigma_0 \left( \frac{\Phi}{\Phi_0}\right)^a .
    \label{andexer}
\end{equation}
Four examples with power law fitted to the data are shown in Fig.\,\ref{fig.twetfit} spanning the whole temperature range.

\begin{figure}
        \includegraphics[width=\columnwidth]{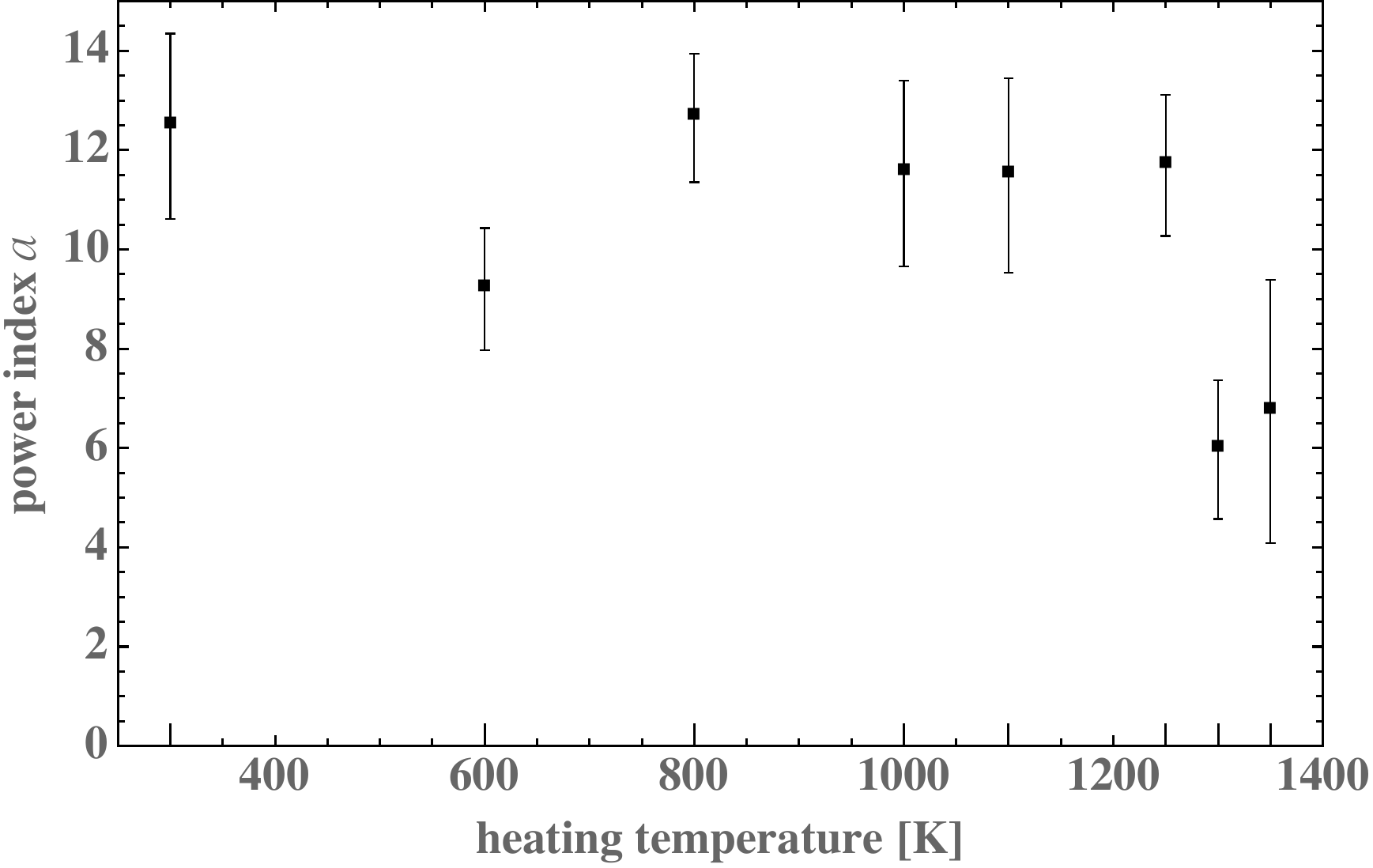}
        \caption{\label{fig.austinpowers} Power indices $a$ (see eq. \ref{andexer}) depending on heating temperature $\mathrm{T_H}$.}
\end{figure}

Because the total dependence is  a power law, we assume that, within the range of data, each of the $\Phi$-dependent factors is a power law in $\Phi$, or
\begin{equation}
N_\Phi = N_0 \left(\frac{\Phi}{\Phi_0}\right)^{a_1}, \,
F_\Phi = F_0 \left(\frac{\Phi}{\Phi_0}\right)^{a_2}, \,
\Phi = \Phi_0 \left(\frac{\Phi}{\Phi_0}\right)^1.
\end{equation}
We then get
\begin{equation}
\sigma = \left(\frac{9}{8 \pi} \right) \left( \frac{1}{d^2} \right) \left(\frac{\Phi}{\Phi_0}\right)^{a_1+a_2+1} N_0 F_0 \Phi_0.
\label{rumpf4}
\end{equation}
In this equation, $N_0$ and $F_0$ are the number of neighbours and average contact force at a certain volume filling factor $\Phi_0$ that has to be chosen arbitrarily and should be in the range of filling factors studied. 
As eq. \ref{rumpf4} holds for all filling factors (within the data range) the power index is
written as\begin{equation}
a = 1 + a_1 + a_2.
\label{powers}
\end{equation}
Fig.\,\ref{fig.austinpowers} shows all fitted power indices $a$ for the different temperatures at which the material was tempered at.

Up to about 1250\,K, the power indices do not vary significantly within the uncertainties.
This is plausible if the grain sizes and shapes do not change significantly, which is consistent with the particle sizes measured. 
Then the arrangement of grains would always be similar at a given filling factor.

For the two highest temperatures at 1300\,K and 1350\,K with tensile strength measurements, the power index is significantly lower. 
Therefore, the number of contacts or the force itself changes in a different way than for the lower temperature data.

In addition to these considerations on power indices, eq. \ref{rumpf4} and \ref{andexer} also yield
\begin{equation}
\sigma_0 = \left(\frac{9}{8 \pi} \right) \left( \frac{1}{d^2} \right)  N_0 F_0 \Phi_0.
\end{equation}
As we are interested in forces, the force is
written as\begin{equation}
F_0 = \frac{8 \pi \sigma_0 d^2}{9 N_0 \Phi_0} .
\label{forcefinal}
\end{equation}
The force strongly depends on grain size, which we did set kind of arbitrarily by grinding. As size $d$ we take the experimentally determined volume-averaged grain size. Knowing the force, a size independent quantity that is still better suited for
applications is the effective surface energy.

\subsection{Surface energy}

We have to keep in mind that systematic subtleties go along with irregular grains and especially that the reduced radii at the contacts are important. However, having said that, we can still use a sticking force as given in eq. \ref{jkr} to define an effective surface energy $\gamma_e$ or 
\begin{equation}
F_0 = 3\pi\gamma_e (d/4).
\label{surftheforce}
\end{equation}
We note that \citet{Steinpilz2019}, based on findings by \citet{Omura2017}, used a sliding force instead, but that was for monodisperse spherical grains of well-known diameters. 
As we have irregular grains in this case, we used the original JKR equation (eq. \ref{surftheforce}), which together with eq. \ref{forcefinal} gives the effective surface energy as
\begin{equation}
    \gamma_e = 1.2 \cdot d \cdot \frac{\sigma_0}{N_0 \Phi_0}.
    \label{gammedef}
\end{equation}
The average coordination number $N_0$ has to be between 2 (chains) and 12 (hexagonal close packing), although for irregular grains variations are possible.  At the lowest filling factor measured, $N_0=5$ seems appropriate \citep{Rumpf1970}. At the highest filling factor, $N_0=8$ seems reasonable. These are just plausible values however \citep{Omura2017,Steinpilz2019}. Considering that some variation  would be possible this adds some systematic uncertainties.
In this case we used the  mean filling factor $\Phi_0=0.44$ of the measured range and the mean coordindation number $N_0=6.5$. Putting in these numbers for the measured data at room temperature, the effective surface energy is $\gamma_e = 0.07 \, \rm J/m^2$ under ambient conditions. We emphasise that this is an effective surface energy, defined by eq. \ref{gammedef} applicable to our problem. We therefore refrain from attributing an error to this value. 

\begin{figure}
        \includegraphics[width=\columnwidth]{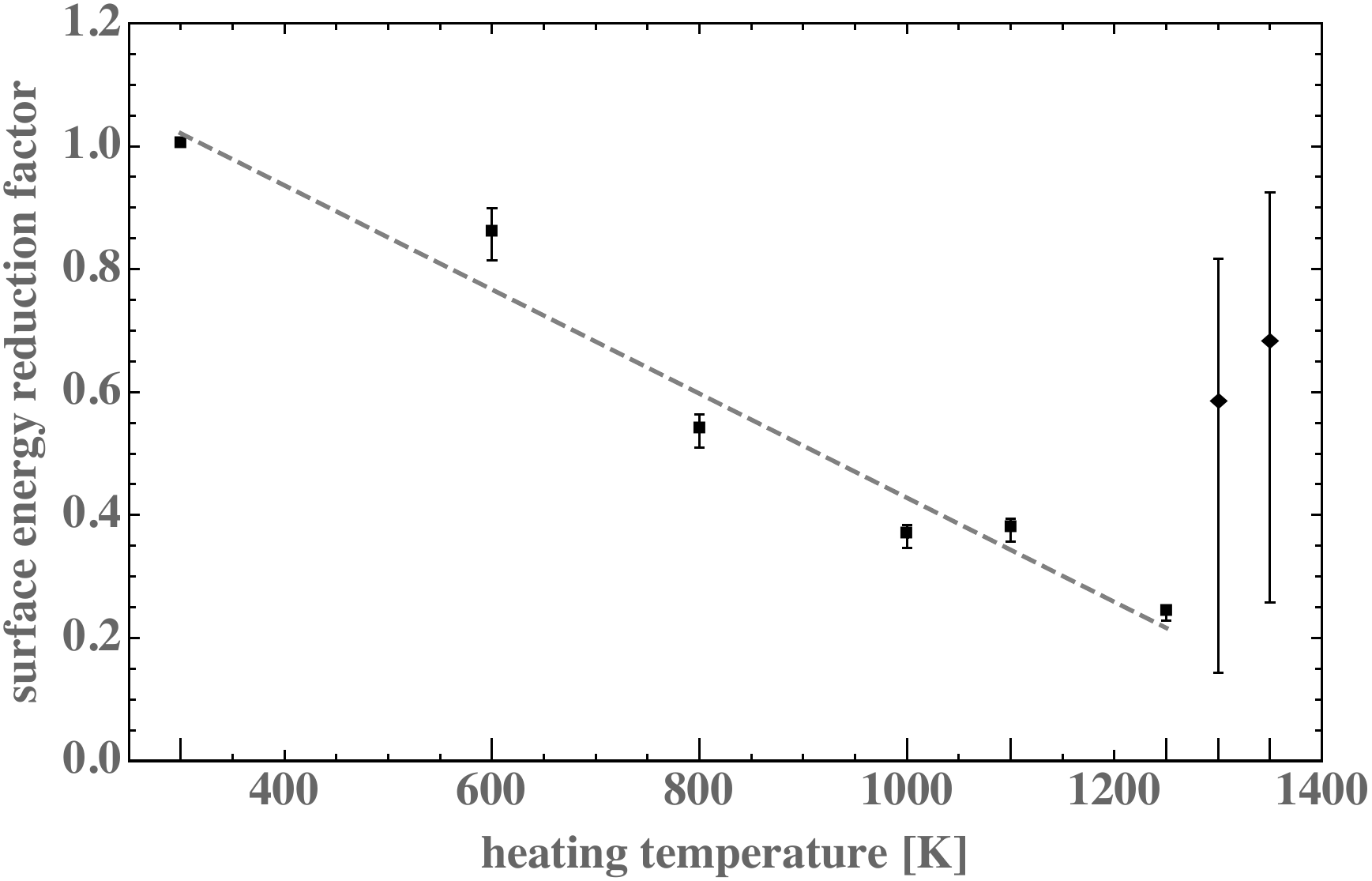}
        \caption{\label{fig.allgammas} Ratio of the surface energy at temperature $T$ to that at room temperature ($\gamma_e(T)/\gamma_e(300 K)$). Typical error bars are of the order of 5\%. The line is a linear trend fitted to the data up to 1250\,K. Error bars for the two highest temperatures are due to volume filling factor dependent differences in contact force or effective surface energy.}
\end{figure}

Independent of the absolute values, the surface energies for different temperatures can be set in relation to each other. In this context, however the power indices of the volume filling factor dependences are important. For varying power index, the choice of $\Phi_0$ matters strongly as $\sigma_0$ would vary in different ways. However, Fig.\,\ref{fig.austinpowers} shows that the power indices do not significantly vary up to 1250\,K. Also the grain sizes of the samples (see above) only start to increase significantly beyond 1250\,K. The data up to 1250\,K are therefore consistent with a constant power and constant grain size. To remove the $\Phi$-dependence in the comparison, we therefore took a constant power index as a plausible assumption and fitted the tensile strength measurements again, but now with the average power index for temperatures up to 1250\,K being fixed to the same value for all these temperatures. We also used the same average grain size of $9.7\,\rm \mu m$ in eq. \ref{gammedef}. Up to 1250\,K the ratios of the surface energies are now independent of the choice of $\Phi_0$ and are shown in Fig.\,\ref{fig.allgammas}. The reference for all ratios is room temperature. 

For the two highest temperatures shown at 1300\,K and 1350\,K, we used both the measured power index and the measured average grain size. However, in this case, because of the different powers of the tensile strength on filling factor, it now matters at which filling factor the surface energies are compared to each other. This is not an artefact of some scaling. The underlying mechanism for this dependence, as outlined above, is a real dependence of the average sticking force on filling factor. Irregular grains just do not have one given sticking force. And as grains of irregular shape are compressed into an aggregate, the specific contacts change. Up to 1250\,K this dependence is the same for all samples. Beyond that temperature, as the increase in average grain size and the change in power indices suggest, the morphology of the grains might change as well, which translates to a variation of the effective surface energy with volume filling factor.

It might be noted, that even though up to 1250\,K, the filling factor dependence of $\gamma_e$ was removed from the ratio of the effective surface energies, there is still a dependence of $\gamma_e$ on $\Phi$. The index $a_2$ related to this cannot be deduced from the data set. However, if the variation in the high temperature values would be taken as guidance, $\gamma_e$ might vary by a factor of 2 for different filling factors. 

Anyway, in general, keeping the volume filling factor constant, the effective surface energy decreases by a factor of about five from room temperature to 1250\,K or to $\gamma_e = 0.014\,\rm J/m^2$. Compared to literature values for crystalline silicates this is about two orders of magnitude less \citep{Parks1984}. However, we note that this is only an effective value. A reduction of two orders of magnitude has been suggested by \citet{Kimura2014} resulting from asperities providing only smaller contacts. The absolute value would be in rough agreement to this idea, keeping in mind that the compositional changes then add another factor to this.
We added a linear trend to the data as we lack a more detailed model. 
For 1400\,K no stable aggregates could be generated owing to the large olivines growing (see grain sizes above).


\subsection{Composition of the untreated material probed by Mössbauer spectroscopy}

The evolution of the effective surface energy with temperature has to be tied to the compositional changes we trace by Mössbauer spectroscopy. For the spectra of the unheated sample and a detailed analysis see appendix\,\ref{appendix1}.

All Mössbauer parameters for the spectrum measured at $\mathrm{293 \, K}$ are listed in table \ref{tab:parameters} and no relevant concentration of additional mineral phases was indicated by X-ray diffraction (XRD). As the relative spectral area $A$ of a mineral approximately represents the amount of iron in this phase relative to the total amount of iron in a material, it is possible to estimate the percentage of weight for all minerals in the sample. This is analysed and discussed in more detail in appendix\,\ref{appendix1}. Furthermore the dependence of the SPM crystallites on environmental temperature delivers insight into their grain sizes, which is discussed in section \ref{subsec:grain_sizes}.

\begin{table*}
    \centering
    \def\arraystretch{1.5}
    \begin{tabular}{ccccc}
    \hline
        \textbf{Mineral} & $\mathrm{\delta \, [mm \, s^{-1}]}$ & $2\epsilon / \Delta E_\mathrm{Q} \, [\mathrm{mm \, s^{-1}]}$ & $B_\mathrm{{HF}}$ \, [T] & $A$ \, [\%]\\ \hline \hline
        \textbf{Olivine} & $\mathrm{1.16\, \pm 0.00}$ & $\mathrm{2.94 \, \pm 0.00}$ & $\mathrm{-}$ & $\mathrm{36.8 \, \pm 0.4}$ \\ \hline
        \textbf{Pyroxene} & $\mathrm{1.16 \, \pm 0.01}$ & $\mathrm{2.06 \, \pm 0.01}$ & $\mathrm{-}$ & $\mathrm{12.2 \, \pm 0.1}$ \\ \hline
        $\mathbf{Fe_3O_4}$ \textbf{A-/B-site} & $\mathrm{0.36 \, \pm 0.05}$ & 0 & $\mathrm{49.1 \, \pm 1.0}$ & $\mathrm{11.1 \, \pm 0.2}$ \\ \hline
        \textbf{SPM} $\mathbf{Fe_3O_4}$ & $\mathrm{0.37 \, \pm 0.01}$ & $\mathrm{0.88 \, \pm 0.02}$ & $\mathrm{-}$ & $\mathrm{34.9 \, \pm 0.4}$ \\ \hline
        $\mathbf{\alpha}$\textbf{-(Fe,Ni)} & $\mathrm{0.02 \, \pm 0.03}$ & 0 & $\mathrm{33.3 \, \pm 0.2}$ & $\mathrm{4.9 \, \pm 0.1}$ \\ \hline
    \end{tabular}
    \caption{Mössbauer parameters of all iron-bearing phases identified in the spectrum of the unheated sample measured at room temperature: isomer shift $\mathrm{\delta}$ relative to $\mathrm{\alpha}$-Fe at room temperature; nuclear quadrupole level shift $\mathrm{2\epsilon}$ (sextets), respectively, quadrupole splitting $\Delta E_\mathrm{Q}$ (doublets); hyperfine magnetic field $B_\mathrm{{HF}}$; and relative spectral area $A$. Values without an error were fixed during the fitting procedure.}
    \label{tab:parameters}
\end{table*}

\subsection{Composition of the tempered material probed by Mössbauer spectroscopy}

Mössbauer spectra were analysed in this manner for a total of ten powder samples heated at different temperatures. Spectra for the samples tempered at the highest temperature ($\mathrm{1400 \, K}$) are shown and discussed in appendix\,\ref{appendix2}. It is noticeable that an exposure to high temperatures changes the composition of chondritic material considerably. 

\begin{figure}
        \includegraphics[width=\columnwidth]{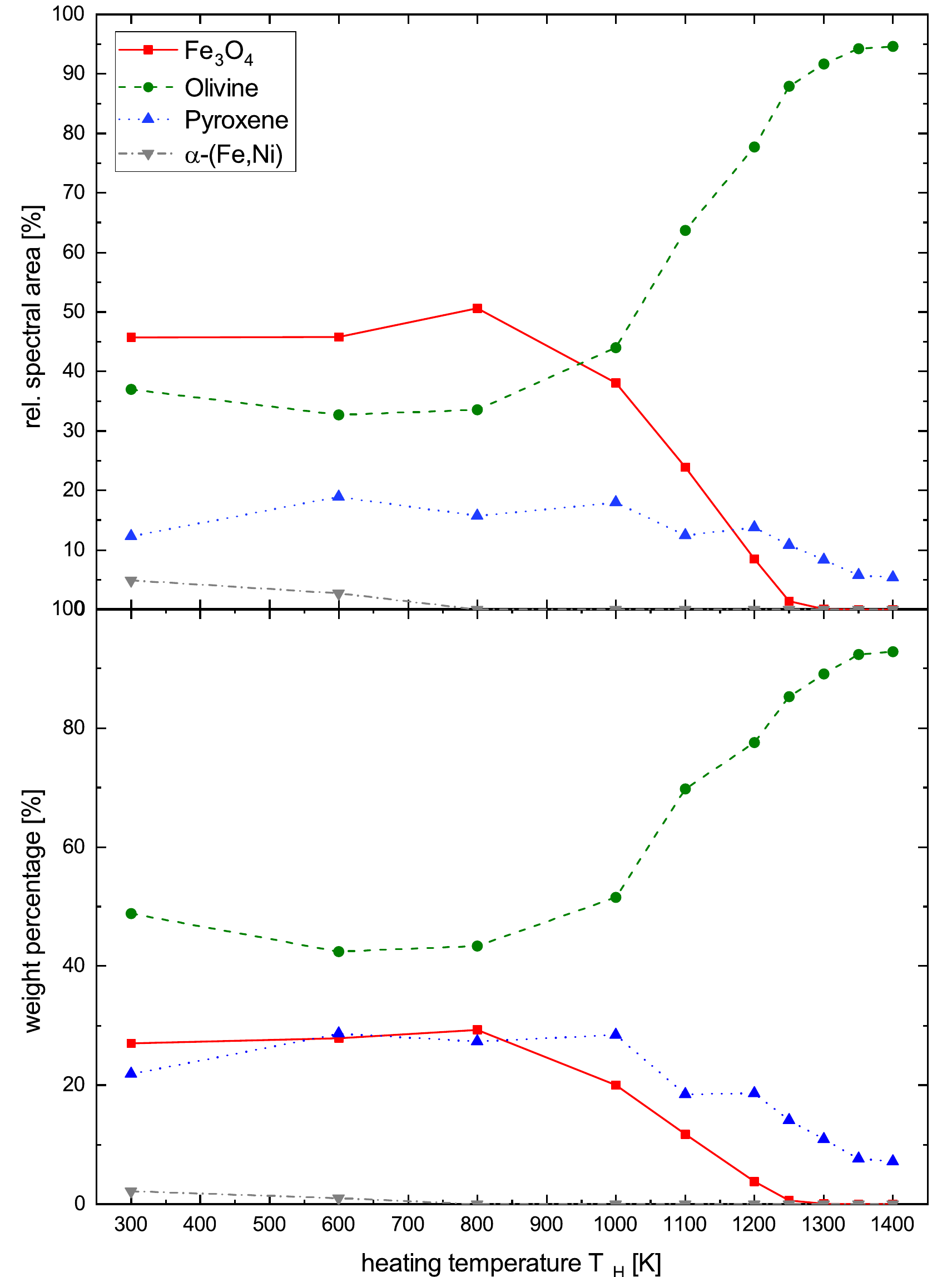}
        \caption{Top panel: Relative spectral area of the contribution of each mineral to the fit. Bottom panel: Weight percentages of all minerals obtained through their relative spectral areas.}
        \label{fig:rel_spectral_areas}
\end{figure}

The dependence of the relative spectral areas, A, of the individual minerals on the temperature they were exposed to prior to the Mössbauer analysis is depicted in the top panel of Fig.\,\ref{fig:rel_spectral_areas}. In this figure the values of $A$ for every heating temperature $T_\mathrm{H}$ were each extracted from their room temperature spectra. While the amount of $\mathrm{Fe_3O_4}$, olivine, and pyroxene is roughly constant for moderate tempering, the sextet structure of kamacite can only be detected for temperatures up to $\mathrm{600 \, K}$. After heating at $\mathrm{800 \, K}$, no $\mathrm{\alpha-(Fe,Ni)}$ is left in the chondritic sample. Exceeding $T_\mathrm{H}=1000 \, \mathrm{K}$ the amount of $\mathrm{Fe_3O_4}$ begins to decrease with every tempering step, until it vanishes completely after exposure to a heating temperature of $\mathrm{1300 \, K}$. This is accompanied by an increase of the spectral area of olivine. After exposures to higher temperatures the types of included minerals, olivines, and pyroxenes no longer change, but the spectra become more dominated by the olivine doublet. We have to keep in mind that the relative spectral area of a mineral is approximately proportional to the number of iron atoms in the sample structuring this exact mineral. Therefore a phase with a higher iron content but the same weight percentage as another, iron-poorer phase in the sample results in a greater relative spectral area. For this reason, all results in Fig.\,\ref{fig:rel_spectral_areas} have been converted into weight percentages based on the following assumption on stoichiometries: $\mathrm{Fe_3O_4}$ for iron oxide, $\mathrm{Fe_{0.95}Ni_{0.05}}$ for kamacite, $\mathrm{MgFeSiO_4}$ for olivine, and $\mathrm{MgFeSi_2O_6}$ for pyroxene. With these values, the dust bulk density $\rho_\mathrm{Dust}(T_\mathrm{H})$ could be estimated as depicted in Fig.\,\ref{fig.bulkdensity}.\newline

\subsection{Influence of tempering on nm-grain sizes}
\label{subsec:grain_sizes}

Besides overall size variations measured by light scattering as quantified above, the Mössbauer data also suggest that iron oxide crystallites become larger. Superparamagnetism occurs when the magnetic anisotropy energy of crystallites $E_\mathrm{A}=K_\mathrm{eff} \, V$, where $K_\mathrm{{eff}}$ is the effective anisotropy constant of the crystallite and $V$ its volume, is comparable to the thermal energy $E_\mathrm{Th}=k_\mathrm{B} \, T$. Decreasing the particle volume allows for thermal fluctuations of its magnetically coupled spin moment, leading to a net paramagnetic behaviour while still maintaining magnetic order. By comparing the areas of SPM to FM magnetite in the Mössbauer spectra, we get an insight into the influence of tempering on $\mathrm{Fe_3O_4}$ crystallite sizes.

\begin{figure}
        \includegraphics[width=\columnwidth]{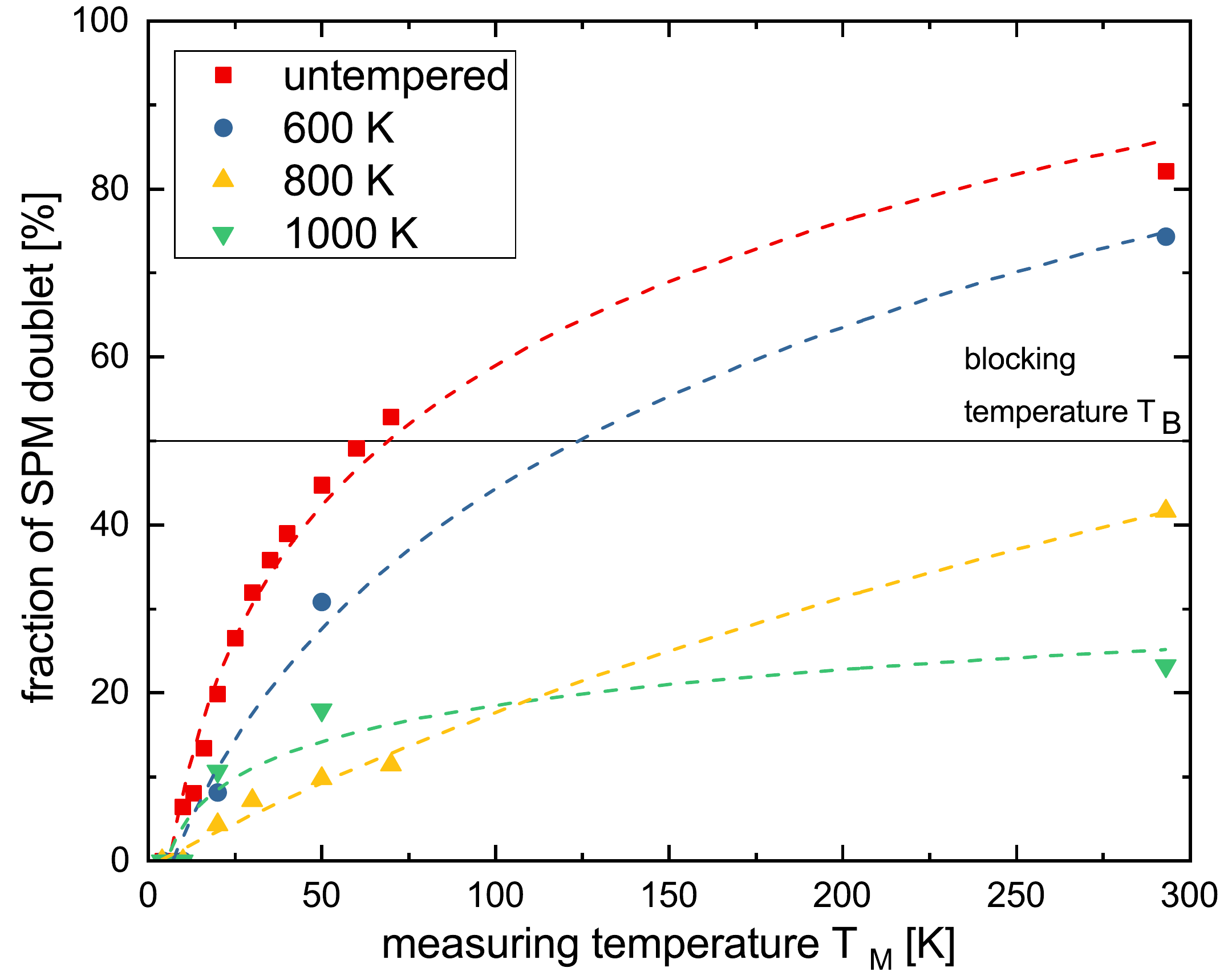}
        \caption{Fraction of SPM $\mathrm{Fe_3O_4}$ doublet area compared to the total $\mathrm{Fe_3O_4}$ contribution for the untreated dust sample and three different heating temperatures $T_\mathrm{H}$. The dashed lines serve as a guide to the eye.}
        \label{fig:blocking_temperatures}
\end{figure}

Fig.\,\ref{fig:blocking_temperatures} depicts the fraction of the SPM magnetite doublet area in relation to the total area assigned to magnetite (SPM doublet plus FM sextets) extracted from Mössbauer spectra. The blocking temperature $T_\mathrm{B}$ is defined as the temperature, where the SPM doublet area equals the sextet area, or speaking in terms of energy, where the magnetic anisotropy energy $E_\mathrm{A}$ approaches the thermal energy $E_\mathrm{{Th}}$. It becomes clear that a growing crystallite volume would result in an increased blocking temperature and could explain the experimental observation. For the untreated dust sample, $T_\mathrm{B}$ is around $\mathrm{60 \, K}$, while a heating temperature of $\mathrm{600 \, K}$ already raises $T_\mathrm{B}$ to approximately $\mathrm{125 \, K}$. The trend of increasing blocking temperatures continues with every tempering event until no more magnetite can be detected in the chondritic powder at approximately $\mathrm{1250 \, K}$. This indicates a growth of nanophase $\mathrm{Fe_3O_4}$ as well as possibly an agglomeration of multiple individual adjacent crystals. Using isolated magnetite nanoparticles as a reference, the increase of $T_\mathrm{B}$ suggests a growth beginning at approximately $\mathrm{4 - 5 \, nm}$ to  $\mathrm{15 - 20 \, nm}$ for powders heated up to $\mathrm{800 \, K}$ (\citet{Landers2016, Goya2003, Haggstrom2008}). 
As particle size in general, but also the size of asperities on surfaces, is important for sticking forces, this influences the effective surface energy.

\section{Conclusions}

We followed the evolution that chondritic material might go through if it drifts inwards in protoplanetary discs. We tempered chondritic dust, milled from a meteorite, at increasing temperatures under vacuum. For each tempered sample, we measured its (ferric) composition by Mössbauer spectroscopy, determined the grain size, and measured the tensile strength of dust aggregates to answer why and how the sticking properties change in warm parts of protoplanetary discs.

Initially four iron-bearing minerals are present, namely metallic iron or kamacite, iron oxides, olivines, and pyroxenes; we note that no additional minerals could be identified by XRD. With increasing temperature the metallic iron already vanishes at moderate temperatures of about 700\,K. Iron oxide grains initially grow in size on the nanometer scale. But the mineral starts to decrease in abundance at about 800\,K, while the relative amount of olivine increases, which could be explained by iron being integrated into the olivines. The iron oxides are completely vanished at about 1250\,K. At the same time, the abundance of olivines levels off. Pyroxene, which is also initially present, starts to decrease in abundance at about 1000\,K, yet some pyroxenes are still present at 1400\,K.

These transformations are related to the evolution in sticking properties. The effective surface energy under ambient conditions at room temperature is $\gamma_e = 0.07 \, \rm J/m^2$. The surface energy monotonously decreases by about a factor of five up to 1250\,K. The high initial effective surface energy and the decrease are related to the iron oxide. The final effective surface energy of $\gamma_e = 0.014 \, \rm J/m^2$ is dominated by the olivines, keeping in mind that the effective value is reduced by about two orders of magnitude compared to literature values for crystalline silicates; this decrease is probably due to asperities dominating the contacts.

The average grain size on the micrometer scale stays the same. Thus changes in contact forces due to variations in the `real' surface energy (composition) and surface curvature, for example from nano-scale changes of grain sizes, both enter in the effective surface energy. 

At still higher temperatures in the range of 1300-1400\,K grains microscopically change size.
The effective surface energy now increases but the increase is sensitive to the filling factor of the aggregates. This is due to the mentioned effect that the influence of grain shape and contact curvatures are traced by the effective surface energy. Depending on the compression of an aggregate these contacts change. While the effective surface energy increases at the high temperatures, aggregates are still weaker owing to the larger grain size. At 1400\,K there is a huge increase in grain size. The sample mass is dominated by grains of about 142\,$\rm \mu m$ diameter then. Stable aggregates can no longer be formed.  

This indicates that at temperatures higher than 1300\,K collisional growth in a sample of chondritic composition is not likely or other mechanisms take over, that is collisions of very sticky sand-size grains close to the melting point of olivine \citep{Bogdan2019}. Melting, or at low pressures, sublimation on longer timescales and recondensation might be an option to form large bodies or only large pebbles. But this would be more of a snowline analogue, which is outside the focus of this paper. Consistent with our findings, \citet{Morbidelli2020} just published a work, suggesting that first planetesimals formed right at this high temperature snowline.

For the lower temperatures, where this is not an issue yet and collisional growth would be the growth mode of choice, we note that we measured under ambient conditions; that is we did not yet take into account that humidity influences surface energies strongly \citep{Steinpilz2019}. 
Subsequent studies on this must complement the data presented in this paper.

With this in mind, it is not yet possible to give a complete picture on particle aggregation in the inner disc. However, based on the subset of data presented,
the strong decrease in surface energy indicates that aggregates grow smaller at higher temperature. The influence of water does likely not alter this general trend. The details will be discussed in a second paper.

If the size of aggregates determines if planetesimal formation is seeded, for example by drag instabilities, the likelihood for planetesimal formation decreases with decreasing distance (or increasing temperature) towards the star with an aggregation limit below 1300\,K and a hard snowline limit at 1400\,K, where other effects might or might not favour formation again.

\section*{Acknowledgments}

This project is supported by DFG grants WU 321/18-1, WE 2623/19-1 and WE 2623/7-1. 
We would also like to thank the anonymous reviewer.




\bibliographystyle{aa}
\bibliography{bibbi} 




%
%

\appendix
        \section{Composition of the untreated material probed by Mössbauer spectroscopy} 
        \label{appendix1}
        
\begin{figure}
      	\includegraphics[width=\columnwidth]{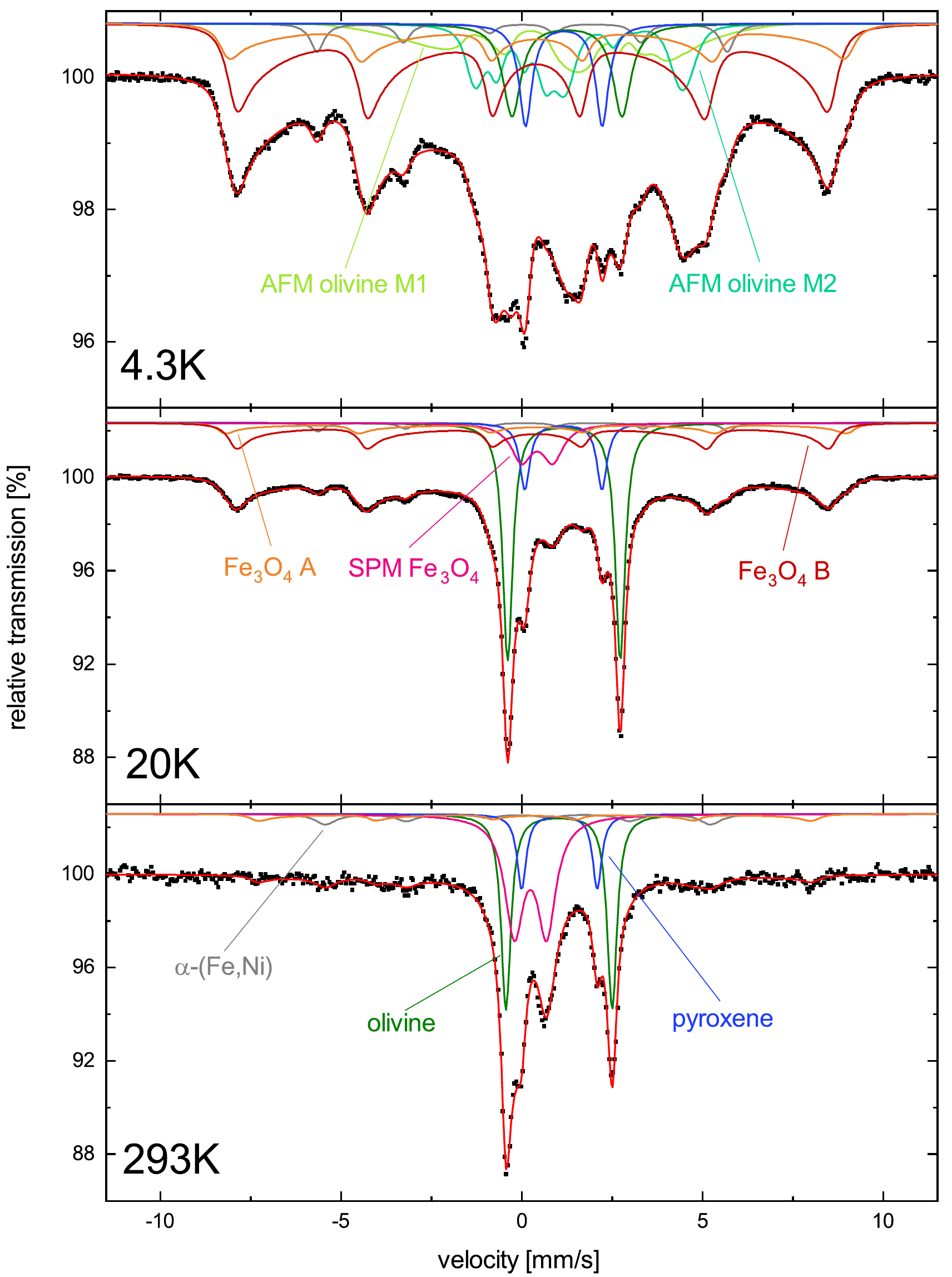}
      	\caption{Mössbauer spectra of the unheated sample measured at $\mathrm{4.3 \, K}$ (top), $\mathrm{20 \, K}$ (middle), and room temperature (bottom). Contributions from various iron-bearing phases are tagged.}
      	\label{fig:M1_ungeheizt}
      \end{figure}

        Fig.\,\ref{fig:M1_ungeheizt} shows the evolution of Mössbauer spectra for the untreated sample measured at various temperatures from $\mathrm{4.3 \, K}$ to $\mathrm{293 \, K}$. 
        At room temperature (bottom panel) the spectrum is dominated by three doublet subspectra reflecting the SPM behaviour of individual iron-bearing phases. Iron-bearing olivines and pyroxenes show no magnetic ordering at room temperature, resulting in a paramagnetic doublet structure in Mössbauer spectra.
        With an isomer shift $\mathrm{\delta}$ of about $\mathrm{1.15 \, mm/s}$ the green and blue doublet can be assigned to members of the olivine ($\mathrm{[Mg_xFe_{1-x}]_2SiO_4}$) and pyroxene group ($\mathrm{[Mg_xFe_{1-x}]_2Si_2O_6}$), respectively. Quadrupole splittings $\mathrm{\Delta E_Q}$ of approximately $\mathrm{2.9 \, mm/s}$ and $\mathrm{2.0 \, mm/s}$, respectively, indicating a 2+ valence for the probed Fe-sites support these assignments (\citet{Oshtrakh2008}, \citet{Klima2007}, \citet{Dyar2006}, \citet{Sklute2006}). 
        
        The remaining pink doublet exhibits Mössbauer parameters which are typically found in iron oxide $\mathrm{Fe_xO_y}$ nanoparticles. Sufficiently small nanoparticles show a  SPM behaviour, that is thermally excited relaxation of the net magnetic moments between easy magnetic directions, leading to the appearance of a doublet structure. Larger magnetic particles might still cause a sextet structure, which can indeed be found in the spectrum (orange). Its isomer shift of approximately $\mathrm{0.35 \, mm/s}$, indicating a contribution of $\mathrm{Fe^{3+}}$ and a hyperfine field of roughly $\mathrm{49 \, T}$, suggests either ferrimagnetic magnetite $\mathrm{Fe_3O_4}$ or maghemite $\mathrm{\gamma-Fe_2O_3}$ (\citet{Lyubutin2009}, \citet{Kuzmann2003}). These iron oxides can normally be identified by the occurrence of two sextet structures caused by $\mathrm{Fe^{3+}}$ on the A- or B-site, respectively. As a consequence of the low amount of magnetically blocked iron oxide in the sample, those two lines cannot be resolved in the room temperature spectrum. Low enough temperatures, however, cause SPM nanoparticles to show magnetically ordered behaviour, which leads to an increase in the spectral area of the sextet and the possibility to resolve two sextet subspectra. An XRD study (not shown here) points to the existence of magnetite rather than maghemite in the sample.
        
        With a rather small relative area a final mineral is discernible by a sextet structure (grey). Its hyperfine field of about $\mathrm{33 \, T}$ and negligible isomer and quadrupole level shift suggest the assignment to $\mathrm{\alpha-Fe}$ (\citet{Preston1962}). A mineral often found in meteorites showing $\mathrm{\alpha-Fe}$-structure is kamacite $\mathrm{\alpha-(Fe,Ni)}$ (\citet{Grandjean1998}, \citet{Cadogan2012}, \citet{Zhiganova2007}, \citet{Menzies2005}), where the fraction of nickel is lower than $\mathrm{10 \, \%}$. As there are no other subspectra visible, the untreated powder shows no iron-bearing phases other than olivine, pyroxene, magnetite, and kamacite. 
        
        By lowering the temperature, magnetic phase transitions can be observed. At $\mathrm{20 \, K}$ (Fig.\,\ref{fig:M1_ungeheizt}, middle panel) the sample temperature falls below the blocking temperature of some of the nanophase $\mathrm{Fe_3O_4}$, resulting in a partial conversion of the SPM doublet to the ferrimagnetic asymmetric sextet structure. The relative spectral area of the sextet increases so that the single A- and B-site subspectra can be resolved (orange and dark red, respectively). The olivine, pyroxene, and kamacite subspectra are not affected that strongly by this drop in temperature.

        Sufficiently low temperatures lead to spontaneous ordering of iron magnetic moments in iron-rich olivines. By cross checking the determined transition temperature with literature values for Néel temperatures of olivines, it is possible to estimate the iron content in this mineralic phase on which the ordering temperature strongly depends. After cooling the powder down to liquid helium temperature, the Mössbauer spectrum shows some dramatic changes (Fig.\,\ref{fig:M1_ungeheizt}, top panel). Iron in olivines is positioned on two non-equivalent sites, namely the M1- and M2-site (\citet{DeOliveria1991}). The magnetic moments of iron on these two positions exhibit an anti-ferromagnetic (AFM) order, corresponding to two complex octet structures (lime green and dark green, respectively). Owing to the distribution of iron contents in the olivines existing in the sample, the lines of these octets are broadened and therefore reproduced using a distribution of hyperfine magnetic fields. Fits based on the Mössbauer parameters for AFM olivines by \citet{Lottermoser1995} lead to a good agreement to the measurement. Furthermore, the paramagnetic olivine doublet decreases in spectral area. A detailed study of this magnetic phase transition, measuring numerous Mössbauer spectra in a temperature range between $\mathrm{4.3 \, K}$ and $\mathrm{20 \, K}$, shows that the olivines contained in the sample have a maximum Néel temperature $T_\mathrm{N}=18 \, \mathrm{K}$. As the Néel temperature of olivine group members increases the more iron they contain, the Fe content of $\mathrm{[Mg_xFe_{1-x}]_2SiO_4}$ present can be roughly estimated to intermediate values of $\mathrm{x \approx 0.5}$ (\citet{Hoye1972}, \citet{Belley2009}).
        
        Aside from that, the increase of magnetically blocked iron oxide is visible to the point where no SPM doublet can be identified in the Mössbauer spectrum. At $\mathrm{4.3 \, K}$, the thermal energy does not suffice to overcome the magnetic anisotropy of nanophase magnetite.

        \section{Composition of the tempered material probed by Mössbauer spectroscopy}
        \label{appendix2}

\begin{figure}
        \includegraphics[width=\columnwidth]{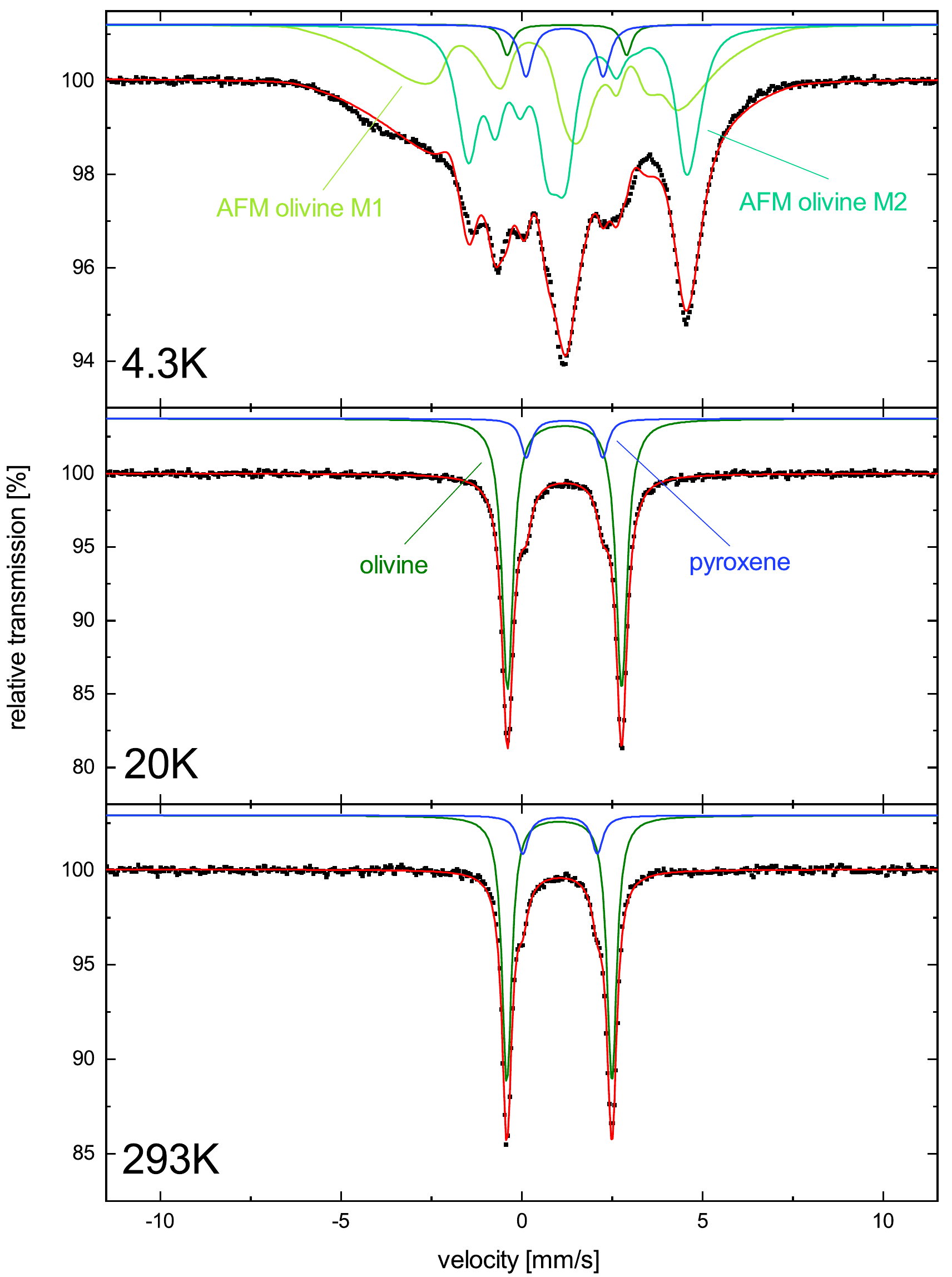}
        \caption{Mössbauer spectra of the sample heated at $\mathrm{1400 \, K}$ measured at $\mathrm{4.3 \, K}$ (top), $\mathrm{20 \, K}$ (middle), and room temperature (bottom). Only iron-bearing silicates are detectable.}
        \label{fig:M1_1400K}
\end{figure}

Fig.\,\ref{fig:M1_1400K} shows the Mössbauer spectra for the samples tempered at 1400\,K.
In comparison to the untreated sample, only iron-bearing silicates are detectable through Mössbauer spectroscopy, recognised by the absence of a $\mathrm{Fe_3O_4}$ SPM doublet and ferrimagnetic sextets as well as by a sextet structure induced by kamacite. Instead, spectra above the Néel temperature $T_\mathrm{N}$ of olivine are composed only of two paramagnetic doublets representing olivine and pyroxene (bottom and middle panels). At $\mathrm{4.3 \, K}$ (Fig.\,\ref{fig:M1_1400K}, top panel) a big fraction of olivine has transitioned to the AFM state. Comparing the fraction of paramagnetic to anti-ferromagnetic olivine for the different heating temperatures, a decrease in paramagnetic olivine at liquid helium temperature with higher $T_\mathrm{H}$ is observable (not shown here). As the Néel temperature $T_\mathrm{N}$ increases with x in $\mathrm{[Mg_{1-x}Fe_x]_2SiO_4}$, this could indicate an enrichment of olivines present with iron and could possibly be connected to the absence of magnetite in the heated samples.


\label{lastpage}
\end{document}